\documentclass[12pt,]{article}
 \usepackage{graphicx}

\begin{document}

 \centerline {\bf \Large The Gould Belt, the de Vaucouleurs–Dolidze Belt, }
  \centerline {\bf \Large  and the Orion Arm}
 \bigskip
 \centerline
 {
 V. V. Bobylev$^{1,2}$
 \footnote [0]{e-mail: vbobylev@gao.spb.ru},
 A. T. Bajkova$^1$
 }
 \bigskip
 \centerline {\small $^{1}$\it Pulkovo Astronomical Observatory, Russian Academy of Sciences,}
 \centerline {\small $^{2}$\it Sobolev Astronomical Institute, St. Petersburg State University,}
 \centerline {\small \it Universitetskii pr. 28, Petrodvorets, 198504 Russia}
 \bigskip

{\noindent Based on masers with measured trigonometric parallaxes,
we have redetermined the spatial orientation parameters of the
Local (Orion) arm. Using 23 sources (the Gould Belt objects were
excluded), we have found that their spatial distribution can be
approximated by a very narrow ellipsoid elongated in the direction
$L_1=77.1\pm2.9^\circ$ whose symmetry plane is inclined to the
Galactic plane at an angle of $5.6\pm0.2^\circ$. The longitude of
the ascending node of the symmetry plane is
$l_\Omega=70\pm3^\circ$. A new estimate for the pitch angle of the
Local spiral arm has been obtained by an independent method:
$i=12.9\pm2.9^\circ$. Previously, a belt of young B stars, the de
Vaucouleurs.Dolidze belt, was pointed out on the celestial sphere
with parameters close to such an orientation. We have refined the
spatial orientation parameters of this belt based on a homogeneous
sample of protostars. The de Vaucouleurs.Dolidze belt can be
identified with the Local arm, with the belt proper as a
continuous band on the celestial sphere like the Gould Belt being
absent due to the peculiarities of the spatial orientation of the
Local arm. Using the entire sample of 119 Galactic masers, we have
shown that the third axis of their position ellipsoid has no
deviation from the direction to the Galactic pole:
$B_3=89.7\pm0.1^\circ.$
  }

  \bigskip

  \noindent{\bf DOI:} 10.1134/S1063773714120020

 \bigskip\noindent


\newpage
 \section*{INTRODUCTION}
A number of structures consisting of young objects are known in
the solar neighborhood: the Gould Belt, de Vaucouleurs--Dolidze
belt, the local system of stars, and the Local (Orion) spiral arm.
The Gould Belt is the nearest and, hence, best-studied object
(Frogel and Stothers 1977; Efremov 1989; P$\ddot{o}$ppel 1997;
Torra et al. 2000). Highly accurate data on the parallaxes and
velocities of stars are being gradually accumulated. They allow
the spatial characteristics of increasingly distant structures to
be refined.

According to present-day estimates, the Gould Belt is a fairly
flat system with semiaxes of $350\times250\times50$ pc, with the
direction of its semimajor axis being near $l=40^\circ$. Its
symmetry plane has an inclination to the Galactic plane of about
$18^\circ$. The longitude of the ascending node is
$l_\Omega=280^\circ$. The Sun is at a distance of $\sim$40 pc from
the line of nodes. The system's center lies at a heliocentric
distance of $\sim$150 pc in the second Galactic quadrant. The
estimate of the direction to the center $l_0$ depends on the
sample age from $130^\circ$ to $180^\circ$. The spatial
distribution of stars is highly nonuniform: a noticeable drop in
density is observed within $\approx$80 pc of the center, i.e., the
entire system has the shape of a doughnut. The well-known open
cluster $\alpha$ Per with an age of $\sim$35 Myr lies near the
center of this doughnut. The system of nearby OB associations (de
Zeeuw et al. 1999) and open star clusters (Bobylev 2006) belongs
to the Gould Belt; a giant neutral hydrogen cloud called the
Lindblad Ring (Lindblad 1967, 2000) is associated with it.

As was pointed out by P$\ddot{o}$ppel (2001), it can be said with
confidence that only relatively nearby stars with a spectral type
no later than B2.5 belong to the Gould Belt. However, the
Hipparcos trigonometric parallaxes with a relative error of less
than 10\% allow only a small solar neighborhood with a radius of
$\sim$120 pc to be analyzed.

The parameters of the de Vaucouleurs.Dolidze belt are known very
approximately. According to Figs. 1 and 2 from Dolidze (1980), the
inclination of its symmetry plane to the Galactic plane ranges
from $16^\circ$ (from young B stars) to $44^\circ$ (from supernova
remnants). Since the distances to the objects were not known, with
the exception of a few $\beta$~Cep stars, the parameters were
determined exclusively from the distribution of stars on the
celestial sphere. This belt reaches the greatest elevation above
the Galactic plane in a direction $l\approx120^\circ$. Dolidze
called this belt the de Vaucouleurs belt and considered it to be
part of the Local spiral arm, a structure as close to the Sun as
the Gould Belt. De Vaucouleurs (1954) believed that the stars he
identified belonged to the bridge between the Galaxy and the
Magellanic Clouds, the Magellanic Stream. We now know that he was
wrong, because the Magellanic Stream consists of hydrogen clouds
and no stars have been detected in it.

Olano (2001) identified the local system of stars with the Local
(Orion) arm and investigated the evolution of this structure with
a mass of $\sim 2\times10^7 M_\odot$ over the last 100 Myr by
numerical simulations. According to Olano's model, there was a
high initial velocity ($\approx$50 km s$^{-1}$) of the gas from
which the local system was formed. It is suggested that such a
velocity could be reached as a result of the interaction with the
Carina--Sagittarius arm. The collision of the gas cloud with the
spiral density wave led to its fragmentation. In this model, such
clusters as the Hyades, the Pleiades, Coma Berenices, and the
Sirius cluster, are considered as the debris of a once single
complex, while contraction of the central regions of the parent
cloud gave rise to the Gould Belt.

As analysis of various data shows, the large-scale spiral pattern
in the Galaxy can be described either by the two-armed logarithmic
spiral model with a constant pitch angle of about $6^\circ$ or by
the four-armed model with a pitch angle of $12-13^\circ$ (Efremov
2011; Vall L ee 2014; Bobylev and Bajkova 2014; Hou and Han
2014), with many authors giving preference to the four-armed
model. In their recent paper, Hou and Han (2014) provide arguments
for a slightly more complex model: in their opinion, the
four-armed model with a variable pitch angle (a polynomial--
logarithmic model) is in even better agreement with the available
data. The Orion arm (Efremov 2011; Xu et al. 2013) is not a a
fully-fledged spiral arm. It is a spur (a small branch) of either
the Carina--Sagittarius arm or the Perseus arm (Hou and Han 2014).
In this paper, we hold this viewpoint.

\begin{figure*}
 \begin{center}
 \includegraphics[width=140mm]{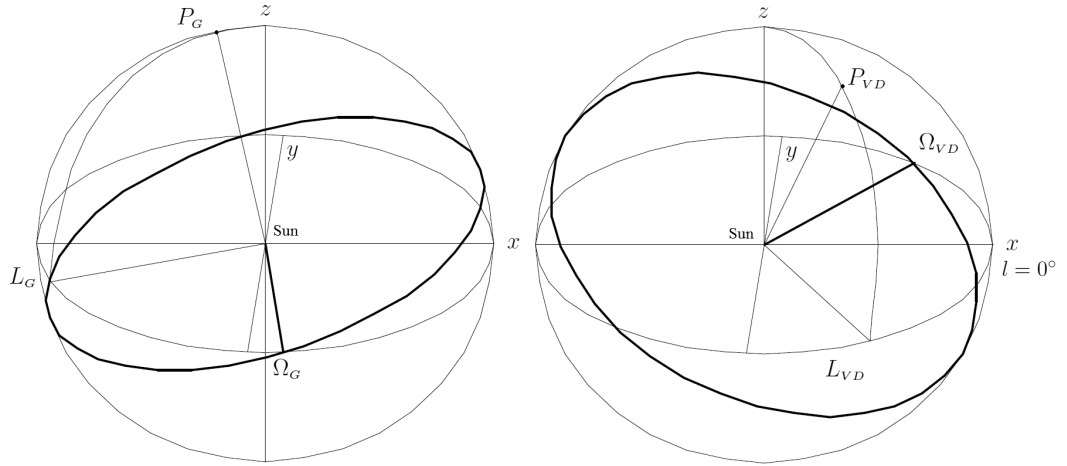}
 \caption{
Orientation of the great circles on the celestial sphere
associated with the Gould Belt (on the left) and the de
Vaucouleurs--Dolidze belt (on the right) relative to the
rectangular Galactic coordinate system.
 }
  \label{f1}
 \end{center}
 \end{figure*}

The orientation of the great circles on the celestial sphere
associated with the Gould Belt and the de Vaucouleurs--Dolidze
belt is schematically shown in Fig.~\ref{f1}. The coordinates of
the pole of the great circle $L,B$ and the longitude of the
ascending node $\Omega$ are marked on each graph.

At present, the masers with their trigonometric parallaxes
measured by VLBI (Xu et al. 2013; Reid et al. 2014) provide an
unprecedented possibility for studying objects in the Local spiral
arm. The relative error in determining the parallaxes by this
method is, on average, less than 10\%. There are about ten such
masers in the Gould Belt, which is not very many. However, their
number in the Local spiral arm is already 37, which allows a
thorough three-dimensional analysis of this system to be
performed. The goal of this paper is to redetermine the spatial
orientation parameters of the Local arm using data on the
positions of masers.

 \section*{THE METHOD}\label{method}
We apply the well-known method of determining the symmetry plane
of a stellar system with respect to the principal (in our case,
Galactic) coordinate system. The basics of the approach were
outlined by Polak (1935); its description can be found in the book
by Trumpler and Weaver (1953). The technique for estimating the
errors in the angles is presented in Parenago (1951) and
Pavlovskaya (1971). Recently, this method was applied to determine
the orientation parameters of the Cepheid system in the Galaxy
(Bobylev 2013).

In the rectangular coordinate system centered on the Sun, the $x$
axis is directed toward the Galactic center, the $y$ axis is in
the direction of Galactic rotation $(l$=$90^\circ$,
$b$=$0^\circ$), and the $z$ axis is directed toward the North
Galactic Pole. Then,
 \begin{equation}
  \begin{array}{lll}
  x=r\cos l\cos b,\\
  y=r\sin l\cos b,\\
  z=r\sin b.
   \label{ff-1}
  \end{array}
 \end{equation}
Let m, n, and k be the direction cosines of the pole of the
sought-for great circle from the $x, y,$ and $z$ axes. The
sought-for symmetry plane of the stellar system is then determined
as the plane for which the sum of the squares of the heights,
$h=mx+ny+kz,$ is at a minimum:
 \begin{equation}
 \sum h^2=\hbox {min}.
 \label{ff-2}
 \end{equation}
The sumof the squares
 \begin{equation}
 h^2=x^2m^2+y^2n^2+z^2k^2+2yznk+2xzkm+2xymn
 \label{ff-3}
 \end{equation}
can be designated as $2P=\sum h^2.$ As a result, the problem is
reduced to searching for the minimum of the function $P$:
 \begin{equation}
 2P=am^2+bn^2+ck^2+2fnk+2ekm+2dmn,
 \label{ff-4}
 \end{equation}
where the second-order moments of the coordinates
 $a=[xx],$
 $b=[yy],$
 $c=[zz],$
 $f=[yz],$
 $e=[xz],$
 $d=[xy],$
written via the Gauss brackets, are the components of a symmetric
tensor:
 \begin{equation}
 \left(\matrix {
  a& d & e\cr
  d& b & f\cr
  e& f & c\cr }\right),
 \label{ff-5}
 \end{equation}
whose eigenvalues $\lambda_{1,2,3}$ are found from the solution of
the secular equation
 \begin{equation}
 \left|\matrix
 {
a-\lambda&          d&        e\cr
       d & b-\lambda &        f\cr
       e &          f&c-\lambda\cr
 }
 \right|=0,
 \label{ff-7}
 \end{equation}
while the directions of the principal axes  $L_{1,2,3}$ and
$B_{1,2,3}$ are found from the relations
 \begin{equation}
 \tan L_{1,2,3}={{ef-(c-\lambda)d}\over {(b-\lambda)(c-\lambda)-f^2}},
 \label{ff-41}
 \end{equation}
 \begin{equation}
 \tan B_{1,2,3}={{(b-\lambda)e-df}\over{f^2-(b-\lambda)(c-\lambda)}}\cos L_{1,2,3}.
 \label{ff-42}
 \end{equation}
The errors in $L_{1,2,3}$ and $B_{1,2,3}$ are estimated according
to the following scheme:
 \begin{equation}
 \varepsilon (L_2)= \varepsilon (L_3)= {{\varepsilon (\overline {xy})}\over{a-b}},
 \label{ff-61}
 \end{equation}
 \begin{equation}
 \varepsilon (B_2)= \varepsilon (\varphi)={{\varepsilon (\overline {xz})}\over{a-c}},
 \label{ff-62}
 \end{equation}
 \begin{equation}
 \varepsilon (B_3)= \varepsilon (\psi)= {{\varepsilon (\overline {yz})}\over{b-c}},
 \label{ff-63}
 \end{equation}
 \begin{equation}
 \varepsilon^2 (L_1)={\varphi^2\cdot \varepsilon^2 (\psi)+\psi^2\cdot \varepsilon^2 (\varphi)
   \over{(\varphi^2+\psi^2)^2}},
 \label{ff-64}
 \end{equation}
 \begin{equation}
 \varepsilon^2 (B_1)= {\sin^2 L_1\cdot \varepsilon^2 (\psi)+\cos^2 L_1\cdot \varepsilon^2 (L_1)
   \over{(\sin^2 L_1+\psi^2)^2}},
 \label{ff-65}
 \end{equation}
where
 \begin{equation}
 \varphi=\cot B_1\cdot \cos L_1, \quad \psi=\cot B_1\cdot \sin L_1,
 \label{ff-66}
 \end{equation}
The three quantities $\overline {x^2y^2}$, $\overline {x^2z^2}$
and $\overline {y^2z^2}$ should be calculated in advance. Then,
 \begin{equation}
 \varepsilon^2 (\overline {xy})= ( \overline{x^2y^2}-d^2)/n,
 \label{ff-71}
 \end{equation}\begin{equation}
 \varepsilon^2 (\overline {xz})= ( \overline {x^2z^2}-e^2)/n,
 \label{ff-72}
 \end{equation}\begin{equation}
 \varepsilon^2 (\overline {yz})= ( \overline {y^2z^2}-f^2)/n,
 \label{ff-73}
 \end{equation}
where $n$ is the number of stars. Thus, the algorithm for solving
the problem consists in (i) setting up the function $2P$
(\ref{ff-4}), (ii) seeking for the roots of the secular equation
(\ref{ff-7}), and (iii) estimating the directions of the principal
axes of the position ellipsoid from Eqs.
(\ref{ff-41})--(\ref{ff-73}). In the classical case, the problem
was solved for a unit sphere ($r=1$), but here we propose to use
the distances that play the role of weights. As can be seen from
Eqs. (\ref{ff-65}), the errors in the directions $L_2$ and $L_3$
coincide; the errors in all the remaining directions are
calculated independently of one another.

\section*{DATA}\label{Data}
\subsection*{\it The Sample of Masers}\label{masewrs}
Based on published data, we gathered information about the
coordinates, line-of-sight velocities, proper motions, and
trigonometric parallaxes of Galactic masers measured by VLBI with
an error, on average, less than 10\%. These masers are associated
with very young objects, protostars of mostly high masses located
in regions of active star formation. The proper motions and
trigonometric parallaxes of the masers are absolute, because they
are determined with respect to extragalactic reference objects
(quasars).

One of the projects to measure the trigonometric parallaxes and
proper motions is the Japanese VERA (VLBI Exploration of Radio
Astrometry) project devoted to the observations of H$_2$O masers
at 22.2 GHz (Hirota et al. 2007) and a number of SiO masers (which
are very few among young objects) at 43 GHz (Kim et al. 2008).

Methanol (CH$_3$OH, 6.7 and 12.2 GHz) and H$_2$O masers are
observed in the USA on VLBA (Reid et al. 2009). Similar
observations are also being carried out within the framework of
the European VLBI network (Rygl et al. 2010), in which three
Russian antennas are involved: Svetloe, Zelenchukskaya, and
Badary. These two programs enter into the BeSSeL project1
\footnote
{http://www3.mpifr-bonn.mpg.de/staff/abrunthaler/BeSSeL/index.shtml}
(Bar and Spiral Structure Legacy Survey, Brunthaler et al. 2011).

The VLBI observations of radio stars in continuum at 8.4 GHz are
being carried out with the same goals (Torres et al. 2009; Dzib et
al. 2011). Radio sources located in the Local arm associated with
young low-mass protostars are observed within the framework of
this program.

Reid et al. (2014) gave a summary of themeasurements of the
trigonometric parallaxes, proper motions, and line-of-sight
velocities for 103 masers. Six more sources of the Local arm from
the list by Xu et al. (2013), EC 95, L 1448C, S1, DoAr21,
SVC13/NGC1333, and IRAS 16293.2422, two red supergiants, PZ Cas
(Kusuno et al. 2013) and IRAS 22480+6002 (Imai et al. 2012), and,
finally, IRAS 22555+6213 (Chibueze et al. 2014), Cyg X-1 (Reid et
al. 2011), IRAS 20143+3634 (Burns et al. 2014) as well as five
low-mass nearby radio stars in Taurus, Hubble 4 and HDE 283572
(Torres et al. 2007), T Tau N (Loinard et al. 2007), HP Tau/G2
(Torres et al. 2009), and V773 Tau (Torres et al. 2012), can be
added to these data; they all belong to the Local arm.

According to Reid et al. (2014), the source G176.51+00.20 with
coordinates $(x,y,z)=(-0.96,0.05,0.0)$ kpc may belong to the Local
arm. However, since it deviates greatly from the general
distribution, we did not include it in the list of masers from the
Local arm. Thus, a total of 37 masers belong to the Local arm. The
entire sample contains 119 sources.

 \begin{figure}
 \begin{center}
 \includegraphics[width=80mm]{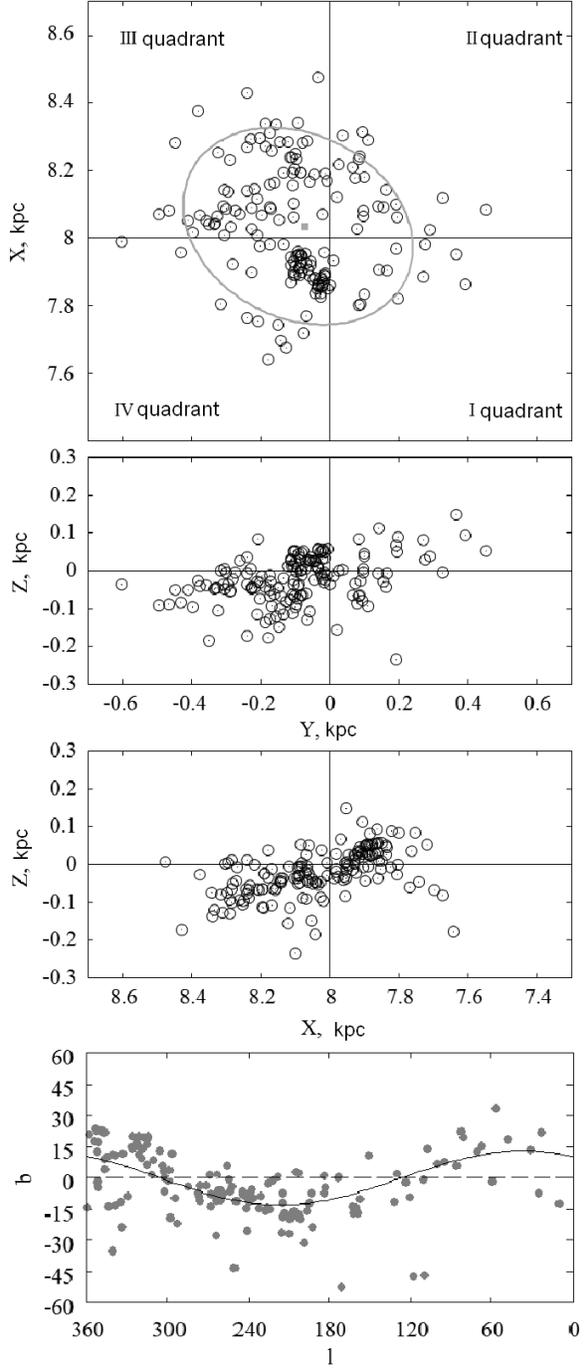}
 \caption{
Positions of 161 O--B2.5 stars in projection onto the Galactic
$XY,YZ$ and $XZ$ planes. The upper panel displays the ellipse
found and the square marks its center; the lower panel presents
the apparent distribution of these stars on the celestial sphere.
 }
  \label{f2}
 \end{center}
 \end{figure}

 \begin{figure}
 \begin{center}
 \includegraphics[width=100mm]{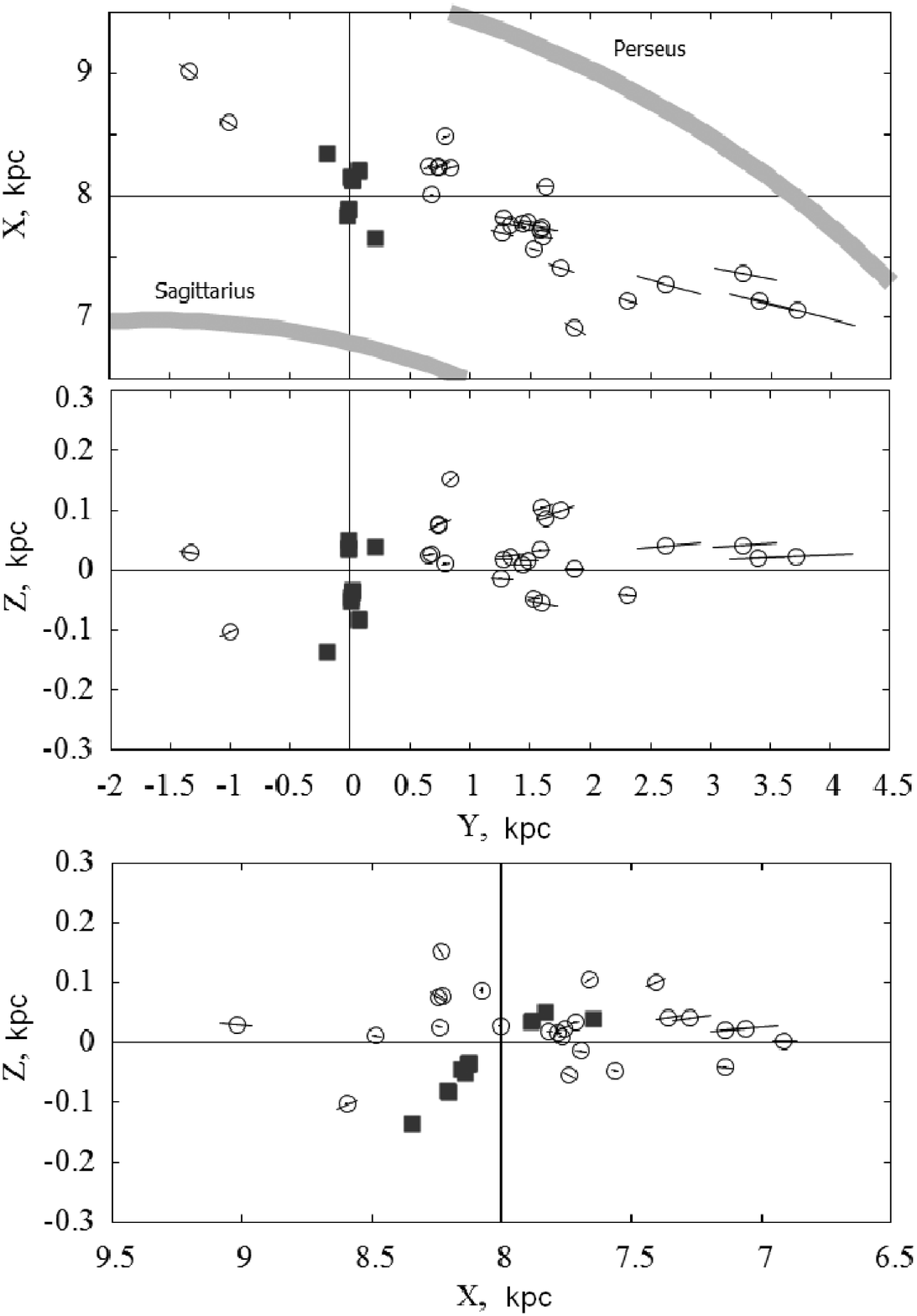}
 \caption{
Positions of Local--arm masers in projection onto the $XY,YZ$ and
$XZ$ planes. The bars denote the distance errors; the Gould-Belt
masers are marked by the dark squares. Fragments of the four-armed
spiral pattern constructed with the pitch angle $i=13^\circ$ are
presented on the upper panel.
 }
  \label{f3}
 \end{center}
 \end{figure}

\subsection*{\it The Sample of O--–B2.5 Stars}\label{OB2stars}
The sample of selected 200 massive (more than 10$M_\odot$) O--B2.5
stars is described in detail in Bobylev and Bajkova (2013). It
contains spectroscopic binary O stars with well-determined
kinematic characteristics within $\sim$3 kpc of the Sun. This
sample also includes 124 stars from the Hipparcos catalogue (van
Leeuwen 2007) with spectral types from B0 to B2.5 whose parallaxes
were determined with a relative error of no less than 10\% and for
which the line-of-sight velocities are known. From this database
we selected 161 stars within 0.7 kpc of the Sun. As was shown by
Bobylev and Bajkova (2013), almost all of them belong to the Gould
Belt.

\section*{RESULTS AND DISCUSSION}
Initially, we tested the method on Gould Belt stars. Based on a
sample of 161 O--B2.5, we found
 \begin{equation}
 \renewcommand{\arraystretch}{1.2}
  \matrix {
  L_1=59\pm12^\circ, & B_1=12\pm2^\circ, \cr
  L_2=148\pm4^\circ, & B_2=-5\pm1^\circ, \cr
  L_3=216\pm4^\circ, & B_3=77\pm1^\circ. \cr
  }
 \label{rezult-11}
 \end{equation}
The principal semiaxes of the ellipsoid in our method are
determined to within a constant and their ratios are
$\lambda_1:\lambda_2:\lambda_3=1:0.78:0.22.$ If the size of the
first semiaxis is taken to be 350 pc, then the ellipsoid will have
sizes, 350 ~ 272 ~ 78 pc, similar to those of the Lindblad Ring.
In contrast to the Lindblad hydrogen Ring whose geometric center
lies in the second quadrant, we obtained slightly different
coordinates of the geometric center of the ``stellar'' ellipsoid:
$x_0=-35\pm13$ pc, $y_0=-92\pm14$ pc, and $z_0=-22\pm 5$ pc.
Obviously, quite a few stars from the Scorpius--Centaurus
association (a dense cloud of points in the fourth quadrant in
Fig. \ref{f2}) exert a great influence on the calculation of these
coordinates. Even if we take the stars from the Scorpius.
Centaurus association with smaller weights, the geometric center
of their distribution will be in the third quadrant. A similar
situation with the geometric center of the Gould Belt determined
from open star clusters was pointed out by Piskunov et al. (2006).
In this case, the parameters of the kinematic center of the Gould
Belt determined from stars are $l_0=160^\circ$ and $R_0=150$ pc
(Bobylev et al. 2004).

On the whole, the parameters found are in good agreement with the
results of determining the orientation of the Gould Belt from
various samples. As can be seen from solution (\ref{rezult-11}),
the orientation of the third axis $L_3,B_3$ is essentially the
same as that in Fig.~\ref{f1}, the inclination to the Galactic
plane is $13\pm1^\circ$, and the longitude of the ascending node
is $l_\Omega=L_3+90^\circ=306\pm4^\circ$. Having analyzed a large
sample of Hipparcos OB stars, Torra et al. (2000) determined the
inclination to the Galactic plane, $16^\circ-22^\circ$, and the
longitude of the ascending node, $l_\Omega=275^\circ-295^\circ$.

The positions of 161 O--B2.5 stars in projection onto the Galactic
$XY,YZ$ and $XZ$ planes as well as their Galactic coordinates $l$
and $b$ are presented in Fig.~\ref{f2}. We denoted the $X,Y$ and
$Z$ axes by big letters, with the $X$ axis being directed away
from the Galactic center. We took the solar Galactocentric
distance to be $R_0=8$ kpc. The sine wave that corresponds to the
parameters (\ref{rezult-11}) found is plotted on the lower panel.
As can be seen from the figure, the stars that we selected by
their spectral type and relative parallax error and that satisfy
the constraints on their coordinates $r\leq0.7$ kpc and $y\leq0.5$
kpc excellently trace the Gould Belt; they are even distributed in
the form of a doughnut. This picture is very similar to the
distribution of neutral hydrogen, the Lindblad Ring. We can
conclude that the method of analysis being applied works and
yields good results. Unfortunately, so far there are only 12
masers and they are concentrated only in six regions associated
specifically with theGould Belt, but the orientation parameters
cannot be found by the described method from such a small number
of sources.

\begin{figure}
 \begin{center}
 \includegraphics[width=100mm]{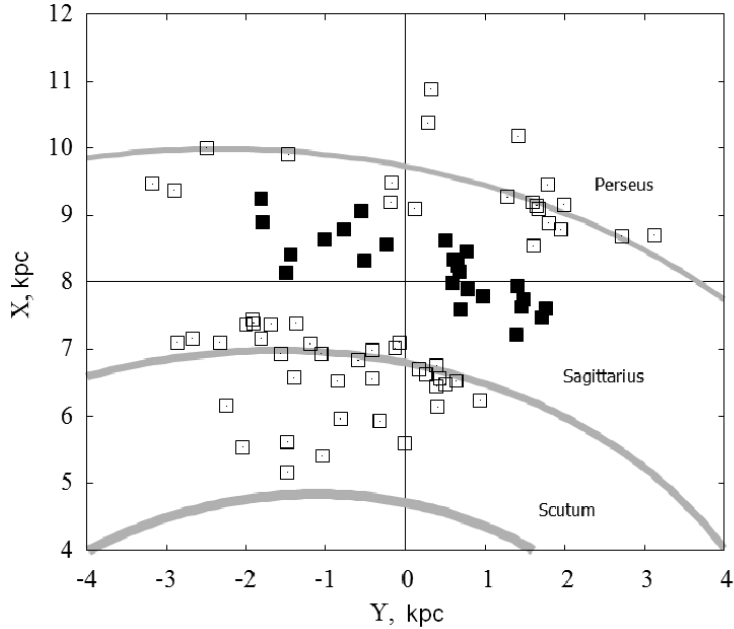}
 \caption{
Positions of OB associations (from Mel'nik and Dambis 2009) in
projection onto the Galactic $XY$ plane. The associations tracing
the Local arm are marked by the dark squares (the Gould-Belt
associations were excluded). }
  \label{f4}
 \end{center}
 \end{figure}

Using 37 masers, we found the following directions of the
principal axes of the position ellipsoid:
 \begin{equation}
 \renewcommand{\arraystretch}{1.2}
  \matrix {
  L_1=~76.0\pm0.2^\circ, & B_1=~~0.7\pm0.0^\circ, \cr
  L_2=166.0\pm2.5^\circ, & B_2=~~1.2\pm0.3^\circ, \cr
  L_3=313.3\pm2.5^\circ, & B_3=88.6\pm0.1^\circ. \cr
  }
 \label{rezult-12}
 \end{equation}
If we remove the Gould Belt masers ($r<0.5$ kpc) and the two stars
from the third quadrant that slightly spoil the solution, then
from the remaining 23 masers of the Local arm we will have
 \begin{equation}
  \matrix {
  L_1=~77.1\pm0.1^\circ, & B_1=~~0.7\pm0.0^\circ, \cr
  L_2=167.1\pm2.9^\circ, & B_2=~~5.6\pm0.1^\circ, \cr
  L_3=340.0\pm2.9^\circ, & B_3= 84.4\pm0.2^\circ. \cr
  }
 \label{rezult-22}
 \end{equation}
The derived ellipsoid is highly elongated; it has the axis ratios
$\lambda_1:\lambda_2:\lambda_3=1:0.15:0.02.$ If, for example, the
half-width of the Local arm is taken, according to the estimate by
Reid et al. (2014), to be 0.33 kpc (this is the size of the second
semiaxis $\lambda_2$), then the ellipsoid found will have the
sizes $2.15\times0.33\times0.05$ kpc. If we take into account the
fact that the geometric center of the Local arm is shifted toward
the Galactic anticenter by $\sim$0.5 kpc, then the derived
ellipsoid is a pretty good approximation of the distribution of
masers in Fig.~\ref{f3} and OB associations of the Local arm in
Fig.~\ref{f4}. The longitude of the ascending node of the symmetry
plane is $l_\Omega=70\pm3^\circ$.

The pitch angle of the spiral pattern is easily estimated via the
direction of the first axis $L_1$,
$i=90^\circ-L_1=12.9\pm2.9^\circ$. Here, when estimating the error
in the angle $i$, we take the largest of the errors in the
longitudes. Solution (\ref{rezult-12}) obtained from all sources
of the Local arm is also suitable for finding the pitch angle,
then $i=14.0\pm2.5^\circ$. Both these values are in good agreement
with the estimate from Xu et al. (2013), where it was determined
from the data on 30 masers in the Local arm, $i=10.1\pm2.7^\circ.$
Note once again that we consider the Local arm to be a spur.
However, the value of $i=12.9\pm2.9^\circ$ found shows that the
Local arm is parallel to two spiral arm segments belonging to the
large-scale structure of the Galaxy, between the Perseus and
Carina. Sagittarius arms. In our opinion, such a position suggests
that the Local arm was formed under the action of the Galactic
spiral density wave. Note that the pitch angle found by Hou and
Han (2014) for the Local arm, $i=1^\circ-3^\circ$, differs
significantly from our result as well as from the result of Xu et
al. (2013).

Figure ~\ref{f3} presents the positions of Local-arm masers in
projection onto the $XY,YZ,$ and $XZ$ planes. To construct the
figure, we took the solar Galactocentric distance $R_0=8$ kpc; the
positions of three segments of the Perseus and Carina--Sagittarius
spiral arms are plotted for this value based on the data from
Bobylev and Bajkova (2014). An inclination of the Gould Belt to
the $X$ axis of about  $+14^\circ$ and an inclination of the more
distant Local arm masers of about $-6^\circ$ (this projection
resembles Fig. ~\ref{f1}) are clearly seen in the distribution of
masers on the $XZ$ plane (the lower panel in Fig. ~\ref{f3}).

Finally, from all the available 119 Galactic masers we find
 \begin{equation}
  \matrix {
  L_1=~34.5\pm0.0^\circ, & B_1=~~0.1\pm0.0^\circ, \cr
  L_2=124.5\pm6.9^\circ, & B_2=~~0.3\pm0.0^\circ, \cr
  L_3=295.5\pm6.9^\circ, & B_3= 89.7\pm0.1^\circ. \cr
  }
 \label{rezult-all}
 \end{equation}
We see that, to within the determination error, the third axis
$B_3$ is directed exactly to the Galactic pole. The nonzero value
of the first axis $L_1$ only suggests that the distribution of
masers in the Galactic $XY$ plane is asymmetric (the empty fourth
quadrant) due to the absence of observations from the Earth's
Southern Hemisphere.

Consider the question of whether the de Vaucouleurs--Dolidze belt
exists like the Gould Belt or this is something else. For this
purpose, it is first necessary to look at the distribution of
Local-arm objects on the celestial sphere. As can be seen from
Fig.~\ref{f3}, there are no masers in the third Galactic quadrant.
To fill this gap, we invoked the data on well-known OB
associations with reliable distance estimates. These associations
are shown in Fig.~\ref{f4}. To construct this figure, we used the
data of Table 1 from Mel'nik and Dambis (2009), where the
distances reconciled with the Cepheid scale are given. A similar,
in principle, picture can also be seen in Fig.~10 from Mel'nik and
Efremov (1995), where a slightly different distance scale was
used.

Note that the distribution of associations in the segments of the
Perseus and Carina.Sagittarius spiral arms in Fig.~\ref{f4} agrees
excellently with the four-armed spiral pattern with the pitch
angle $i=13^\circ$ plotted according to Bobylev and Bajkova
(2014). The question about the number of spiral arms in the Galaxy
has not yet been solved completely. A detailed discussion of this
question can be found in a recent paper by Hou and Han (2014), who
used the latest extensive data on traces of the Galactic spiral
structure: neutral and ionized hydrogen, giant molecular clouds,
and masers.

 \begin{figure}
 \begin{center}
 \includegraphics[width=120mm]{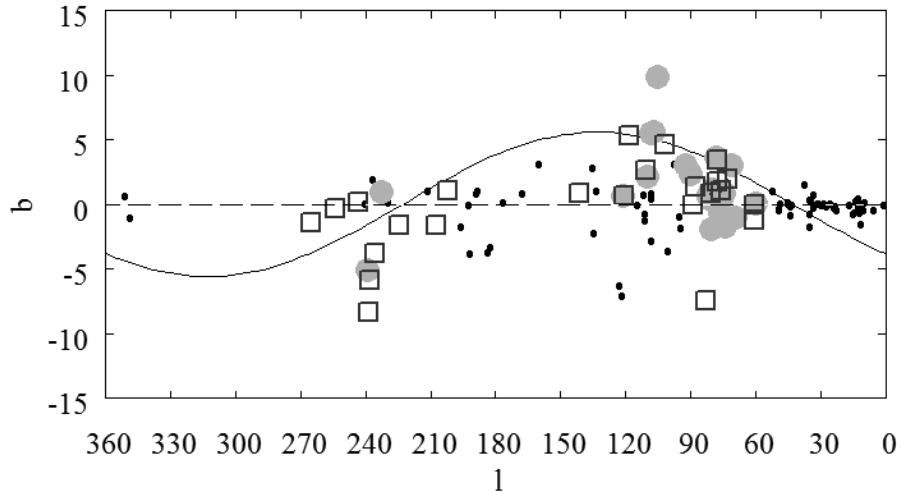}
 \caption{
Apparent positions of masers and OB associations on the celestial
sphere. The Local-arm masers are marked by the big gray circles,
the distant Galactic masers that do not belong to the Local arm
are marked by the filled circles, and the OB associations of the
Local arm are denoted by the open squares. }
  \label{f5}
 \end{center}
 \end{figure}

The apparent distribution of Local-arm masers and OB associations
on the celestial sphere is presented in Fig.~\ref{f5}, all of the
remaining Galactic masers are also shown, and the sine wave
constructed according to solution (\ref{rezult-22}) is plotted. As
can be seen from the figure, the distant masers lie very close to
the Galactic plane toward the Galactic center, but there are many
sources in a direction near $l=80^\circ$ and they have fairly
large elevations. In this direction, one of the OB associations
deviates greatly from the general distribution, possibly because
of the large distance measurement error. This association is
Cyg~OB4 ($r=0.8$~kpc). It is most likely slightly nearer and is
associated with the Gould Belt. All of the remaining associations
in this direction lie very close together, in good agreement with
the positions of Local-arm masers. In a direction near
$l=250^\circ$, the positions of OB associations excellently
complement the overall picture.

It can be seen from Fig.~\ref{f5} that there are two clumps of
Local--arm objects near $l=77^\circ$ and in the diametrically
opposite direction $l=257^\circ$. The remaining space along the
plotted sine wave is empty. Therefore, the de Vaucouleurs.Dolidze
belt proper as a continuous band on the celestial sphere like the
Gould Belt is simply absent. There is the Local arm. Its spatial
location relative to the Sun is such that we observe its two
projections as two dense clumps of stars in the directions
$l=77^\circ$ and $l=257^\circ$. Moreover, when we separate the
Gould Belt as an independent structure, we just cannot see quite a
few Local-arm masers or OB associations in the direction of the
second axis of the Local arm $l=167-347^\circ$, given the
narrowness of the Local arm, because there are none of them.

It is important to note the following. The deviation of the
system's symmetry plane from the Galactic plane means that some
forces cause the stars to rise/sink in the vertical direction.
Studying the velocities of stars is of great interest in this
regard. Such vertical oscillations are known for the velocities of
Gould-Belt stars. There is even the term "vertical oscillation
axis" (Comer$\acute{o}$n 1999; Lindblad 2000). It is interesting
to check the velocities of Local-arm stars, especially those in
the first and second quadrants, where the stars reach considerable
heights (Fig.~\ref{f4}). However, this is the objective of our
next paper.

\section*{CONCLUSIONS}\label{conclusions}
Based on published data, we gathered information about the
Galactic masers with their trigonometric parallaxes measured by
VLBI. The sources located in the Local arm are of greatest
interest for this work. To determine the spatial orientation of
this system, we applied the well-known method of three-dimensional
analysis that consists in determining the parameters of the
position ellipsoid.

Initially, the method was tested on Gould-Belt stars. For this
purpose, we used a sample of 161 O--B2.5 stars whose distances are
known with errors of no less than 10--15\%. The determined
orientation parameters of the Gould Belt system were shown to be
in good agreement with the results of other authors.

Using 23 masers (the Gould-Belt objects were excluded) from the
Local arm, we found that their spatial distribution could be
approximated by a very narrow ellipsoid elongated in the direction
$L_1=77.1\pm2.9^\circ$ whose symmetry plane is inclined to the
Galactic plane at an angle of $5.6\pm0.2^\circ$. The longitude of
the ascending node of the symmetry plane is
$l_\Omega=70\pm3^\circ$.

The orientation of the first axis $L_1$ allows a new estimate for
the pitch angle of the spiral pattern to be obtained for the Local
arm, $i=12.9\pm2.9^\circ$. This value shows that the Local arm is
parallel to two spiral arm segments belonging to the large-scale
structure of the Galaxy, between the Perseus and Carina.
Sagittarius arms. In our opinion, this position suggests that the
Local arm was formed under the action of the Galactic spiral
density wave.

Previously, the de Vaucouleurs--Dolidze stellar belt with
parameters close to such an orientation was known on the celestial
sphere. In this paper, we refined significantly the spatial
orientation parameters of this belt based on a very homogeneous
sample of protostars. The results obtained lead to the conclusion
that we can identify the de Vaucouleurs.Dolidze belt with two
dense clumps of stars in the directions $l=77^\circ$ and
$l=257^\circ$ that do not contain any Gould-Belt objects, with the
de Vaucouleurs--Dolidze belt proper as a continuous band on the
celestial sphere like the Gould Belt being simply absent.

The main conclusion of our analysis is as follows. If the Local
arm is considered as a whole, then the third axis of the position
ellipsoid has an insignificant deviation from the Galactic z axis,
as can be seen from solution (\ref{rezult-12}). If, however, the
Gould Belt and the rest of the Local arm are considered
separately, then the position ellipsoid for each of these parts
has a significantly different orientation, solutions
(\ref{rezult-11}) and (\ref{rezult-22}).

Using the entire sample of 119 masers, we showed that the third
axis of their position ellipsoid has no significant deviation from
the direction to the Galactic pole.

\section*{ACKNOWLEDGMENTS}
\bigskip
We are grateful to the referee for the useful remarks that
contributed to an improvement of the paper. This work was
supported by the "Nonstationary Phenomena in Objects of the
Universe" Program P--21 of the Presidium of the Russian Academy of
Sciences.

\bigskip

\bigskip

{\noindent\bf\Large REFERENCES}

\bigskip

\bigskip

{

1. V. V. Bobylev, Astron. Lett. 30, 159 (2004).

2. V. V. Bobylev, Astron. Lett. 32, 816 (2006).

3. V. V. Bobylev and A. T. Bajkova, Astron. Lett. 39, 532 (2013).

4. V. V. Bobylev, Astron. Lett. 39, 753 (2013).

5. V. V. Bobylev and A. T. Bajkova, Mon. Not. R. Astron. Soc. 437,
1549 (2014).

6. A. Brunthaler, M. J. Reid, K. M. Menten, X.- W. Zheng, A.
Bartkiewicz, Y. K. Choi, T. Dame, K. Hachisuka, K. Immer, G.
Moellenbrock, et al., Astron. Nach. 332, 461 (2011).

7. R. A. Burns,  Y. Yamaguchi, T. Handa, T. Omodaka, T. Nagayama,
A.Nakagawa, M.Hayashi, T. Kamezaki, J. O. Chibueze, et al.,
arXiv:1404.5506 (2014).

8. J. O. Chibueze, H. Sakanoue, T. Nagayama, T. Omodaka, T. Handa,
T. Kamezaki, R. Burns, H. Kobayashi, H. Nakanishi, M. Honma, et
al., arXiv:1406.277 (2014).

9. F. Comer . on, Astron. Astrophys. 351, 506 (1999).

10. M. V. Dolidze, Sov. Astron. Lett. 6, 394 (1980).

11. S. Dzib, L. Loinard, L. F. Rodriguez, A. J. Mioduszewski, and
R. M. Torres, Astrophys. J. 733, 71 (2011).

12. Yu.N. Efremov, Sites of Star Formation inGalaxies (Nauka,
Moscow, 1989) [in Russian].

13. Yu. N. Efremov, Astron.Rep. 55, 108 (2011).

14. J. A. Frogel and R. Stothers, Astron. J. 82, 890 (1977).

15. T. Hirota, T. Bushimata, Y. K. Choi, M. Honma, H. Imai, K.
Iwadate, T. Jike, S. Kameno, O. Kameya, R. Kamohara, et al., Publ.
Astron. Soc. Jpn. 59, 897 (2007).

16. L. G. Hou and J. L. Han, arXiv:1407.7331 (2014).

17. H. Imai, N. Sakai, H. Nakanishi, H. Sakanoue, M. Honma, and
T.Miyaji, Publ. Astron. Soc. Jpn. 64, 142 (2012).

18. M. K. Kim, T. Hirota, M. Honma, H. Kobayashi, T. Bushimata, Y.
K. Choi, H. Imai, K. Iwadate, T. Jike, S. Kameno, et al., Publ.
Astron. Soc. Jpn. 60, 991 (2008).

19. K. Kusuno, Y. Asaki, H. Imai, and T. Oyama, Astrophys. J. 774,
107 (2013).

20. F. van Leeuwen, Hipparcos, the New Reduction of the Raw Data
(Springer, Dordrecht, 2007).

21. P. O. Lindblad, Bull. Astron. Inst. Netherland 19, 34 (1967).

22. P. O. Lindblad, Astron. Astrophys. 363, 154 (2000).

23. L. Loinard, R. M. Torres, A. J. Mioduszewski, L. F. Rodriguez,
R. A. Gonzalez-Lopezlira, R. Lachaume, V. Vazquez, and E.
Gonzalez, Astrophys. J. 671, 546 (2007).

24. A.M.Melnik and Yu. N. Efremov, Astron. Lett. 21, 10 (1995).

25. A. M. Melnik, and A. K. Dambis, Mon. Not. R. Astron. Soc. 400,
518 (2009).

26. C. A. Olano, Astron. Astrophys. 121, 295 (2001).

27. P. P. Parenago, Tr. Gos. Astron. Inst. Shternberga 20, 26
(1951).

28. E. D. Pavlovskaya, Practical Works on Stellar Astronomy, Ed.
by P. G. Kulikovskii (Nauka, Moscow, 1971), p. 162 [in Russian].

29. A. E. Piskunov, N. V. Kharchenko, S. R` 'oser, E. Schilbach,
and R.-D. Scholz, Astron. Astrophys. 445, 545 (2006).

30. I. F. Polak, Introduction to Stellar Astronomy (ONTI, Moscow,
Leningrad, 1935) [in Russian].

31. W. G. L. P . oppel, Fundament. Cosm. Phys. 18, 1 (1997).

32. W. G. L. P . oppel, ASP Conf. Ser. 243, 667 (2001).

33. M. J. Reid, K. M. Menten, X. W. Zheng, A. Brunthaler, L.
Moscadelli, Y. Xu, B. Zhang, M. Sato, M. Honma, T. Hirota, et al.,
Astrophys. J. 700, 137 (2009).

34. M. J. Reid, J. E. McClintock, R. Narayan, L. Gou, R. A.
Remillard, and J. A. Orosz, Astrophys. J. 742, 83 (2011).

35. M. J. Reid, K. M. Menten, A. Brunthaler, X.W. Zheng, T.M.
Dame, Y. Xu, Y.Wu, B. Zhang, A. Sanna, M. Sato, et al., Astrophys.
J. 783, 130 (2014).

36. K. L. J. Rygl, A. Brunthaler, M. J. Reid, K. M. Menten, H. J.
van Langevelde, and Y. Xu, Astron. Astrophys. 511, A2 (2010).

37. J. Torra, D. Fern . andez, and F. Figueras, Astron. Astrophys.
359, 82 (2000).

38. R. M. Torres, L. Loinard, A. J. Mioduszewski, and L. F.
Rodriguez, Astrophys. J. 671, 1813 (2007).

39. R. M. Torres, L. Loinard, A. J. Mioduszewski, and L. F.
Rodriguez, Astrophys. J. 698, 242 (2009).

40. R. M. Torres, L. Loinard, A. J. Mioduszewski, A. F. Boden, R.
Franco-Hernandez, W. H. T. Vlemmings, and L. F. Rodriguez,
Astrophys. J. 747, 18 (2012).

41. R. J. Trumpler and H. F.Weaver, Statistical Astronomy (Univ.
California Press, Berkely, 1953).

42. J. P. Vall . ee, Mon. Not. R. Astron. Soc. 442, 2993 (2014).

43. G. de Vaucouleurs, Observatory 74, 23 (1954).

44. Y.Xu, J. J. Li,M. J.Reid,K.M.Menten,X.W.Zheng, A. Brunthaler,
L. Moscadelli, T. M. Dame, and B. Zhang, Astrophys. J. 769, 15
(2013).

45. P. T. de Zeeuw, R. Hoogerwerf, J. H. J. de Bruijne, A. G. A.
Brown, and A. Blaauw, Astron. J. 117, 354 (1999).

}

\end{document}